\documentclass[twocolumn,showpacs,preprintnumbers,amsmath,amssymb,prd]{revtex4}

\pdfoutput=1

\usepackage{graphicx}	
\usepackage{dcolumn}	
\usepackage{bm}		
\usepackage{sistyle}		

\DeclareMathOperator \dm {d}

\newcommand{\beq}{\begin{equation}}
\newcommand{\eeq}{\end{equation}}
\numberwithin{equation}{section}

\begin{document}

\preprint{AEI-2008-027}

\title{Blandford's Argument: The Strongest Continuous Gravitational Wave Signal}

\author{Benjamin Knispel}
\email{Benjamin.Knispel@aei.mpg.de}
\author{Bruce Allen}
\email{Bruce.Allen@aei.mpg.de}
\affiliation{Max-Planck-Institut f\"ur Gravitationsphysik (Albert-Einstein-Institut) and Leibniz Universit\"at Hannover\\Callinstr. 38, 30167 Hannover, Germany}

\begin{abstract}
  For a uniform population of neutron stars whose spin-down is
  dominated by the emission of gravitational radiation, an old argument
  of Blandford states that the expected gravitational-wave amplitude
  of the nearest source is independent of the deformation and rotation
  frequency of the objects. Recent work has improved and extended this
  argument to set upper limits on the expected amplitude from neutron
  stars that {\it also} emit electromagnetic radiation. We restate these
  arguments in a more general framework, and simulate the evolution of such a
  population of stars in the gravitational potential of our galaxy. The simulations
  allow us to test the assumptions of Blandford's argument on a realistic model
  of our galaxy. We show that the two key assumptions of the argument
  (two dimensionality of the spatial distribution and a steady-state frequency
  distribution) are in general not fulfilled. The effective scaling dimension $D$
  of the spatial distribution of neutron stars is significantly larger than two, and
  for frequencies detectable by terrestrial instruments the frequency distribution
  is not in a steady state unless the ellipticity is unrealistically large.
  Thus, in the cases of most interest, the maximum expected
  gravitational-wave amplitude {\it does} have a strong dependence on the
  deformation and rotation frequency of the population. The results strengthen
  the previous upper limits on the expected gravitational-wave amplitude from
  neutron stars by a factor of 6 for realistic values of ellipticity.
\end{abstract}


\maketitle

\section{Introduction}

Continuous emission from spinning neutron stars is a promising source
of gravitational waves, but so far no detections have been
reported. This begs the question ``what is the largest expected
amplitude of the continuous signal from nonaxisymmetric neutron
stars?''  There might exist a (so far undetected) population of
spinning neutron stars whose dominant energy loss goes into the
production of gravitational waves, rather than into electromagnetic
radiation.  These are often called ``gravitars''; we show later in
this paper that gravitars set an upper limit on the amplitude of
gravitational waves from spinning neutron stars that are also
emitting electromagnetic radiation.

In 1984 Blandford found a simple analytic relationship between the
expected maximum amplitude of gravitational waves emitted by gravitars
and their average galactic birthrate. This argument was not published
but it is documented by a citation in
[\onlinecite{original_blandford}].  The argument was recently revised
in [\onlinecite{collaboration2006}].  This paper revises both the
original and the revised Blandford arguments, and shows that two key
assumptions of these arguments do not hold in a realistic galactic
model of gravitars.  This paper corrects the assumptions of the
argument, and then investigates how the conclusions are affected by
this change.

We stress that while this paper studies the behavior of a population
of galactic gravitars, it does not make a plausibility case for the
possible existence of such objects, or study their potential
astrophysical implications. The study itself is nevertheless
interesting, because even if gravitars do {\em not} exist, they
provide a relevant upper bound on the gravitational-wave emission
by objects (such as rapidly-spinning neutron stars) that {\em do} exist.

The paper is organized as follows.  Sec.\ \ref{sec:analytic} reviews
Blandford's argument and its assumptions. A simple analytic
calculation is used to derive the frequency-space distribution of the
sources. This allows a sharper statement of the conclusion and
clarifies the dependence upon the assumptions. The aim of this paper
is to test whether these assumptions are fulfilled in a realistic
model of our galaxy and, if the assumptions do not hold, what the
consequences are.  Sec.\ \ref{sec:num_model} describes a numerical
simulation of the Galaxy, and Sec.\ \ref{sec:results} presents
results for the simulated spatial and frequency distribution of
gravitars at the present time. The simulated spatial distributions do
not satisfy the assumptions of Blandford's argument: they are not
two-dimensional and uniform. Sec.\ \ref{bland:revise} uses these
simulated distributions to recompute the expected maximum
gravitational-wave amplitudes from gravitars.  As shown in
[\onlinecite{collaboration2006}], the maximum expected gravitational
wave amplitudes from gravitars are upper limits for the gravitational
wave amplitudes from neutron stars spinning down through combined
electromagnetic and gravitational-wave emission.  Previous work
assumed that all neutron stars are formed with the same (high)
birth-frequency.  Here, the argument is generalized to cover a
continuous distribution of initial frequencies.  This is followed by a
short conclusion.

For realistic models of neutron stars, the general upper limit on
gravitational-wave emission set by considering the gravitar case
applies for gravitational-wave frequencies $f\gtrsim
\SI{250}{Hz}$. The reader who wants to skip all the details and just
see the final result is advised to look at Fig.\ \ref{amplitude}
which is the main result of the paper.

\section{Blandford's argument: an analytic description}\label{sec:analytic}

\subsection{Frequency Evolution and a First Analysis}\label{originbland}

If a rotating neutron star has a nonaxisymmetric shape, it will radiate
away rotational energy by the emission of gravitational waves.
It is straightforward to derive the equations describing the frequency
evolution of gravitars. Their spin-down due to their nonaxisymmetric shape
is given by \beq \dot f= - \frac{32 \pi^4}{5}\frac{G
  I}{c^5}\varepsilon^2 f^5\label{dotf}, \eeq where $f$ is the
frequency of the emitted gravitational waves, which is twice the
spin frequency of the gravitar. $G$ is Newton's gravitational constant,
$I$ is the momentum of inertia with respect to the rotational axis, $c$
is the speed of light and $\varepsilon=\frac{I_1-I_2}{I}$ is the ellipticity
of the gravitar. Integrating (\ref{dotf}) gives the frequency of
gravitational waves emitted at time $t$ as \beq f\left( t\right) =
\left(f_\textsc{b}^{-4} + \beta^{-1}\varepsilon^2t\right)^{-\frac{1}{4}}
\quad\text{with } \beta=\frac{5}{128\pi^4}\frac{c^5}{GI},
\label{foft} \eeq assuming an initial (birth)
gravitational-wave frequency $f_\textsc{b} =
f\left(0\right)$. The constant $\beta$ is approximately
\beq \beta^{\frac{1}{3}} = 5.3 \times
\left(\frac{\SI{e38}{kg\,m^2}}{I}\right)^{\frac{1}{3}} \text{kHz}.
\eeq
We infer from (\ref{foft}) the spin-down timescale
\beq \tau_\textsc{gw} \left(\varepsilon,f\right) := \beta\varepsilon^{-2}f^{-4}
\approx \SI{4.6}{Gyr} \; \biggl(\frac{10^{-7}}{\varepsilon}\biggr)^2
\biggl(\frac{\SI{100}{Hz}}{f} \biggr)^4 \label{sdtimescale},\eeq
which is the time for a gravitar born at a gravitational wave
frequency $f_\textsc{b} \gg f$ with ellipticity $\varepsilon$ to spin
down to gravitational-wave frequency $f$ via the emission of
gravitational waves.  These equations allow one to calculate the
gravitational-wave frequency at the present time for any gravitar
given its birth frequency, ellipticity, and age. This in turn allows
one to determine the frequency distribution of a population of gravitars.

The strain amplitude $h$ of gravitational waves emitted by a gravitar at
distance $r$ to the detector and assuming optimal mutual orientation
(gravitar sky position given by the unit vector orthogonal to the plane
of the detector arms; gravitar spin axis parallel to this vector) is given
by \beq h = 4\pi^2\frac{GI}{c^4}\frac{\varepsilon
  f^2}{r}\label{strain}.  \eeq Given a model for the spatial
distribution of gravitars, this allows one to determine the
distribution of gravitational-wave amplitudes.

Let us begin by giving Blandford's original argument in a more
complete form than the single-paragraph version given in
[\onlinecite{original_blandford}].

Assume there is a population of galactic gravitars, which remain
undetected because they do not emit electromagnetic waves.  Ref.\ 
[\onlinecite{palomba}] (particularly Sec.\ 6 and the appendix) shows
the conditions necessary for a neutron star to be a gravitar.
Ref.\ [\onlinecite{popov}] is a simulation of a population of
isolated neutron stars accreting matter from the interstellar medium
and demonstrates that quite a few neutron stars may in fact meet these
conditions. Taken together, these two papers establish a detailed
plausibility argument for the possible existence of a population of
gravitars.

Assume the neutron stars are uniformly distributed in a thin two dimensional
galactic disk with radius $R$ and assume that the time between
gravitar births in our galaxy is constant: $\tau_\textsc{b}\approx
\SI{30}{yrs}$. Assume that all gravitars are born with the same
ellipticity $\varepsilon$ and high birth frequency $f_\textsc{b}$.
The frequency of each gravitar will then evolve according to
(\ref{foft}).

Consider an interval $\left[f_1,f_2\right]$ of gravitational wave
frequencies with $f_1,f_2 \ll f_\textsc{b}$. Let us consider wide ranges
of frequencies corresponding to the broad-band sensitivity of modern
interferometric detectors.  Then, from (\ref{foft}) the time a gravitar
will spend in this interval of frequencies is 
\beq t_{12} = \left[1 -
  \left(\frac{f_1}{f_2}\right)^4\right]\tau_\textsc{gw}\left(\varepsilon,f_1\right).
\eeq
The number of sources in this frequency interval is given by
$N_{12} = \frac{t_{12}}{\tau_\textsc{b}}$ if $t_{12} \geqslant
\tau_\textsc{b}$ and depends on the choice of the frequency
band. Because of the assumed two dimensionality and uniformity of the
spatial distribution the average distance $r_\text{cl}$ to the closest gravitar
in this range of frequencies can be written as \beq r_\text{cl} =
\frac{R}{\sqrt{N_{12}}} =
R\sqrt{\frac{\tau_\textsc{b}}{\tau_\textsc{gw}\left(\varepsilon,f_1\right)}}\left[1-\left(\frac{f_1}{f_2}\right)^4\right]^{-\frac{1}{2}}.\label{n_in_band}
\eeq
The formula given in Ref.\ [\onlinecite{original_blandford}] agrees
with Eq.\ (\ref{n_in_band}) if one assumes that the factor in
square brackets is of order unity, which is the case for the latest generation of broad-band
interferometric gravitational-wave detectors.

To estimate the gravitational-wave amplitude of the strongest source
requires a bit of care.  To get the flavor of the original argument,
consider a one-octave frequency band $[f, 2f]$.  The quantity in
square brackets in (\ref{n_in_band}) is $15/16$ which we approximate
as unity.  Substituting the distance to the closest source
(\ref{n_in_band}) into (\ref{strain}) and neglecting the fact that
$f^2$ can vary by up to a factor of four within the octave, gives the
amplitude of the strongest source in this one-octave frequency band
to be
\beq
\label{strongestsource}
h = 4\pi^2\frac{GI}{c^4}\frac{\varepsilon f^2 \sqrt{\tau_\textsc{gw}}
}{R \sqrt{\tau_\textsc{b}}} =
4\pi^2\frac{GI}{c^4}\frac{\sqrt{\beta}}{R \sqrt{\tau_\textsc{b}}} 
= \sqrt{\frac{5 G I}{8 c^3 R^2 \tau _\textsc{b}}}.
\eeq
The Blandford argument is simply the observation that this
amplitude is (1) independent of the population's deformation $\varepsilon$ and (2)
independent of frequency $f$.

Blandford's argument may also be stated in terms of a comparison
between two different model Galaxies, each containing a similar
populations of gravitars but each having a different (but constant)
value of the ellipticity.

\subsection{Restating Blandford's Argument}\label{restate:bland}

The previous paragraph is a rigorous version of Blandford's original
argument. We now generalize this, building on the methods first presented
in [\onlinecite{collaboration2006}].

Let us first define useful quantities to describe a population of gravitars.
As before $r$ denotes the distance between gravitar and detector, $f$ is the
frequency of gravitational waves emitted, and $\varepsilon$ is the ellipticity
of the gravitar.

In this section $t$ measures the age of the gravitars. A gravitar with
age $t = 0$ is born at the present time, whereas $t > 0$ for a
gravitar born in the past.

Because optimal mutual orientation of the gravitar's spin and the detector's
normal axis is assumed, the gravitational-wave amplitude depends only on
the distance $r$ but not on the sky position. Therefore, it is useful to define
the probability $\dm\!P_\text{r}$ of finding a gravitar born time $t$ ago at the
present time in a spherically symmetric shell $\left[r,r+\dm\!r\right]$ around the Sun
\footnote{Detector and Sun can be assumed collocated on galactic scale.}.
The probability can be written in terms of a probability density
$\varrho_\text{r}\left(r, t\right)$ as
\beq \dm\!P_\text{r} = \varrho_\text{r}\left(r, t\right) \dm\!r. \eeq

Moreover, let us define the probability $\dm\!P_f$ of finding gravitars born
time $t$ ago with ellipticity $\varepsilon$ in a present-time frequency band
$\left[f,f+\dm\!f\right]$. $\dm\!P_f$ can be written in terms of
a probability density $\varrho_f\left(\varepsilon, f, t\right)$ as
\beq \dm\!P_f = \varrho_f\left(\varepsilon, f, t\right) \dm\!f. \eeq
 
 Note, that the probability densities are normalized by 
 $\int_0^\infty\dm\!r\,\varrho_\text{r}\left(r,t\right) = 1 \; \forall t$ and
 $\int_0^\infty\dm\!f\,\varrho_f\left(\varepsilon,f,t\right) = 1 \; \forall \varepsilon,t$.

For further generalization consider a continuous distribution of gravitational
wave frequencies at birth instead of a single, high value. Let $\dm\!P_{f_0}$
be the probability of the birth frequency being in a band
$\left[f_0, f_0 + \dm\!f_0\right]$. The corresponding probability density
$\varrho_{f_0}\left(f_0\right)$ is defined by
\beq \dm\!P_{f_0} = \varrho_{f_0}\left(f_0\right)\dm\!f_0,\eeq
normalized as before.
Frequency change by redshift from cosmological evolution is neglected since
all gravitars considered are within our galaxy.

To link the initial frequency distribution $\varrho_{f_0}$ to the
present-time distribution $\varrho_f$, consider a gravitar with
ellipticity $\varepsilon$ whose current frequency is $f$, and let
$f_0\left(\varepsilon, f, t\right)$ denote the gravitar's frequency at time $t$
in the past.
Solving (\ref{foft}) for the birth frequency yields
\beq f_0\left(\varepsilon,f,t\right) = \left(f^{-4} - \beta^{-1}\varepsilon^2t\right)^{-\frac{1}{4}}.\eeq
The probability density $\varrho_f$ can be rewritten in terms of the initial frequency
distribution $\varrho_{f_0}$ by a change of variables. The fraction  of gravitars
in a birth frequency band $\left[f_0,f_0+\dm\!f_0\right]$ is the same as the fraction in
a present time frequency band $\left[f,f+\dm\!f\right]$, so the identity
$\varrho_f\dm\!f = \varrho_{f_0}\dm\!f_0$ yields
\begin{align}
\varrho_f\left(\varepsilon, f, t\right)\dm\!f & =  \varrho_{f_0}\left(f_0\left(\varepsilon, f, t\right)\right)\frac{\partial f_0\left(\varepsilon, f, t\right)}{\partial f}\dm\!f \label{f:dist_first}\\
& =  \varrho_{f_0}\left(f_0\left(\varepsilon, f, t\right)\right) \frac{f_0^5\left(\varepsilon, f, t\right)}{f^5}\dm\!f,
\end{align}
from which
\beq \varrho_f\left(\varepsilon, f, t\right) = \varrho_{f_0}\left(f_0\left(\varepsilon, f, t\right)\right) \frac{f_0^5\left(\varepsilon, f, t\right)}{f^5}\label{f:dist}\eeq
immediately follows.

To allow for a time-dependent birthrate of galactic gravitars, let $\dot n\left(t\right)$ be
the birthrate as a function of $t$. The number of gravitars $\dm\!N$
formed during a short time interval $\left[t, t+\dm\!t\right]$ is then
$\dm\!N = \dot n\left(t\right)\dm\!t$.

The number $\dm\!\tilde N $ of gravitars in a thin spherical shell $\left[r,r+\dm\!r\right]$
around the position of the Sun, with frequencies in $\left[f,f+\dm\!f\right]$,
with fixed ellipticity $\varepsilon$, formed in a time interval $\left[t,t+\dm\!t\right]$ ago
is then given by
\begin{eqnarray}
\dm\!\tilde N &=&\dm\!P_\text{r} \times \dm\!P_f\times \dm\!N\nonumber\\
&=&\varrho_\text{r}\left(r,t\right)\dm\!r \times \varrho_f\left(\varepsilon,f,t\right)\dm\!f \times \dot n\left(t\right)\dm\!t.
\end{eqnarray}

From (\ref{strain}) it follows that for fixed $\varepsilon$ and $f$ there is a unique,
invertible mapping $r\left(h\right)$ from the amplitude of gravitational waves
$h$ to the distance $r$ of the gravitar from the Sun. A change of variables from $r$ to $h$ yields
\beq \dm\!\tilde N = \varrho_\text{r}\left(r\left(h\right),t\right)\frac{\dm r\left(h\right)}{\dm\!h}\dm\!h \times
\varrho_f\left(\varepsilon,f,t\right)\dm\!f \times \dot n\left(t\right)\dm\!t \eeq
and the number $M\left(f_1,f_2,h_\text{max}\right)$ of gravitars with a gravitational
wave amplitude $h \geqslant h_\text{max}$ in a frequency band $\left[f_1,f_2\right]$
and ages $t \leqslant \overline{t}$ is given by integration as
\begin{align}
&M\left(f_1,f_2,h_\text{max}\right) =\nonumber\\
&\int_0^{\overline{t}}\!\!\dm\!t\,\dot n\left(t\right) \int_{f_1}^{f_2} \dm\!f\
\varrho_f\left(\varepsilon,f,t\right)\int_{h_\text{max}}^{\infty}\dm\!h\,
\varrho_\text{r}\left(r\left(h\right),t\right)\frac{\dm r\left(h\right)}{\dm\!h}.\label{eq:M}
\end{align}
Here, the integral over $h$ is performed for a fixed $\varepsilon$ (assuming the
same ellipticity for every gravitar) and fixed $f$ and $t$. After integrating out the
dependence on $h$ the follow-up integration over $f$ weights the previous
integral by the frequency density. The last integration sums the distributions
from different birth times weighted by the galactic neutron star birthrate at that time.

Before Eq.\ (\ref{eq:M}) is used to rederive Blandford's result, let us prove
that the frequency distribution from a single birth frequency has reached a steady
state ($\partial_t\varrho_f= 0$) at frequency $f$, if $f\left(t\right) < f < f_\textsc{b}$,
where $f\left(t\right)$ is given by (\ref{foft}). Consider a frequency band
$\left[f,f+\dm\!f\right]$ which is wide enough to contain at least one gravitar at all
times $t$. If $f\left(t\right) < f < f_\textsc{b}$, the constancy of the birthrate guarantees
that if and only if a gravitar leaves the frequency band by the lower boundary
another gravitar will enter the frequency band from higher frequencies.
The assumption of a steady state is crucial. If the distribution has not reached a
steady state in a certain frequency band there will be no sources in that
band and there is no contribution to the integral in (\ref{eq:M}).

Let us now re-derive Blandford's result by using a density function
$\varrho_\text{r}\left(r,t\right) = 2r/R^2$ which describes a
population of galactic gravitars uniformly distributed in a flat
two-dimensional disk with radius $R$ \footnote{This assumes that the
  Sun is farther from the edge of the disk than the closest expected
  source. Since the expected loudest sources are very close to the Sun
  this assumption is justified.}. Further, assume a constant birthrate
$\dot n\left(t\right) = \frac{1}{\tau_\textsc{b}}$ and a single high
birth frequency $f_\textsc{b}$ such that
$\varrho_{f_0}\left(f_0\right) = \delta \left(f_0 -
  f_\textsc{b}\right)$.  Inserting $r\left(h\right)$ into (\ref{eq:M})
by solving (\ref{strain}) for $r$ we find after a slightly technical
but straightforward calculation \beq M\left(f_1,f_2,h_\text{max}\right)
=\frac{5GI}{\tau_\textsc{b}c^3R^2} \int_{f_1}^{f_2}
\frac{\dm\!f}{f}\int_{h_\text{max}}^{\infty}\frac{\dm\!h}{h^3}.
\label{eq:M_general}
 \eeq The
integrations are trivial and yield \beq
M\left(f_1,f_2,h_\text{max}\right) = \frac{5GI}{2\tau_\textsc{b}c^3R^2}
h_\text{max}^{-2}\ln\left(\frac{f_2}{f_1}\right).\label{eq:M_sec} \eeq
Let us follow [\onlinecite{collaboration2006}] and assume a 50\%
chance of detection, corresponding to $M = 1/2$. One finds a maximum
gravitational-wave strain [\onlinecite{collaboration2006}]  \beq h_\text{max} =
\sqrt{\frac{5GI}{\tau_\textsc{b}c^3R^2}\ln\left(\frac{f_2}{f_1}\right)}.\label{eq:M_typical}
\eeq This result can also be directly compared with the earlier result
(\ref{strongestsource}) from the cruder analysis, by setting $f_2 =
2f_1$ and setting $M=1$.  The values of $h_\text{max}$ obtained by these
two different analyses disagree by about 40\%, but are
independent of deformation and frequency $f_1$.

For a broad band search performed today we assume $\ln\left(f_2/f_1\right) \approx 1$.
Then (\ref{eq:M_typical}) gives the largest amplitude expected under the assumptions
\footnote{$R=\SI{10}{kpc}$, $\tau_\textsc{b}=\SI{30}{yrs}$, $I=\SI{e38}{kg\,m^2}$.} 
from galactic gravitars as $h_\text{max} \approx \num{4e-24}$.

For later comparison with the realistic galactic model, let us calculate the dimensionless
averaged fractional frequency density $\hat\varrho_f\left(\varepsilon,f\right)$ in
the population. It is defined by $\dm\!\hat P_f = \hat\varrho_f\left(\varepsilon,f\right)
\frac{\dm\!f}{f}$ being the probability to find gravitars with a fixed ellipticity
$\varepsilon$ in a frequency band $\left[f, f+\dm\!f\right]$,
\beq
\hat\varrho_f\left(\varepsilon,f\right) = \frac{f}{N_\text{tot}}\int_0^{\overline{t}}\!\!\dm\!t\,\dot n\left(t\right) \varrho_f\left(\varepsilon,f,t\right),\label{ffracdens}
\eeq
where $N_\text{tot} = \int_0^{\overline{t}}\!\!\dm\!t\,\dot n$ is the number of gravitars
formed during the timespan $\overline t$. Using the same assumptions as for the
derivation of (\ref{eq:M_sec}) yields
\beq
\hat\varrho_f\left(\varepsilon,f\right) = \frac{4\beta}{\tau_\textsc{b}}\varepsilon^{-2}f^{-4}.\label{ffracdens_simple}
\eeq
For fixed ellipticity the averaged fractional frequency density falls off with
$f^{-4}$ and scales for fixed frequency as $\varepsilon^{-2}$.

Let us summarize the assumptions made for this analytic calculation of the
strongest gravitational-wave signal from galactic gravitars. Assume all
gravitars are born at a single high birth frequency with fixed ellipticity and
constant birthrate, and reside in a two-dimensional, uniform distribution,
i.e.\ in a thin galactic disk. Assume their spin-down is governed by the emission of
gravitational waves as described by Eq.\ (\ref{foft}). Adopting these
assumptions and a 50\% chance of actual detection, the largest amplitude
$h_\text{max}$ of gravitational waves emitted by galactic gravitars in a frequency
band $\left[f_1,f_2\right]$ is given by (\ref{eq:M_typical}). Thus, a precise
statement about $h_\text{max}$ is the following
\begin{quote}
\textsl{\textbf{Result:} Assume the existence of a population
of galactic gravitars with uniform, two-dimensional spatial
distribution, single, high birth frequency, fixed ellipticity
$\varepsilon$, and constant birthrate. Choose a frequency band
$\left[f,sf\right]$ with scale $s > 1$ large enough such that there
is at all times at least one gravitar in this band. Then the
largest amplitude $h_\text{max}$ of gravitational waves
emitted by galactic gravitars in this band is independent of $f$
and $\varepsilon$ and depends only on the scale $s$.}
\end{quote}
Searching wider ranges of frequencies increases the value of $h_\text{max}$
because the absolute number of gravitars in wider ranges of frequency increases.
However, in (\ref{eq:M_typical}) the gain from going to higher frequencies
grows slowly, as the square root of the logarithm, because the gravitars
spend less time at higher frequencies.

\subsection{A Natural Limit to the Result}

Because of the crucial assumption of a steady-state distribution in
frequency there are obvious limits to this simple model. The time to
reach a steady state in a given narrow frequency band $\left[f,f+\dm\!f\right]$
must be at least of the same order of magnitude as the
spin-down timescale (\ref{sdtimescale}), because otherwise no
gravitar will have spun down to frequencies contained in the band.

If all gravitars are born at the same high frequency the time to reach
a steady state is exactly the spin-down age. If on the other hand
there is a continuous distribution of initial frequencies, then
reaching a steady state in a certain frequency band requires longer
evolution times.  Only then most of the gravitars in that band are ones that
have spun down from higher frequencies.  Over time this effect
``washes out'' any effects of the initial frequency distribution.

There is a natural limit to the result due to the finite age of the Universe,
since no gravitar can be older than the Universe itself. An even better limit
would be the age of the Galaxy, or rather that of the galactic neutron star
population. However, since the age of the Universe is known much more
accurately than the age of the Galaxy, and since they differ only by a factor
of order 2, we will use the age of Universe in all of our estimates below.

The age of the Universe $t_0$ can be calculated from Hubble's constant $H_0$
as $t_0 = \frac{2}{3}H_0^{-1}$. Then the finiteness sets limits on the values
of $\varepsilon$ and $f$ for which the population has reached a steady
state.  We easily find from (\ref{sdtimescale}) that the population is in a
steady state for gravitational-wave frequencies that satisfy \beq \varepsilon^2f^4 >
\frac{3}{2}H_0\beta.  \eeq Fixing the ellipticity, one can calculate a
frequency \beq \tilde f\left(\varepsilon\right) =
\left(\frac{3H_0\beta}{2\varepsilon^2}\right)^\frac{1}{4} =
\SI{76}{Hz}\left(\frac{\num{e-7}}{\varepsilon}\right)^\frac{1}{2}\label{eps_f}\eeq
above which the population can be assumed to be close to a steady
state.

We would like to stress that a realistic population with a
continuous distribution of initial frequencies has to have evolved
over a time $T \approx \text{few}\times\tau_\textsc{gw}$ to be in
steady-state.  Thus the true value of $\tilde
f\left(\varepsilon\right)$ is larger by a factor of a few, and falls
into the frequency range of highest sensitivity in modern
interferometric gravitational-wave detectors (between \SI{100}{Hz} and
\SI{300}{Hz}).

The range of ellipticities for which the assumption of steady state breaks down
is then given by
\beq
\tilde \varepsilon\left(f\right) \lesssim \num{5.8e-8} \left(\frac{\SI{100}{Hz}}{f}\right)^2.
\eeq
In general it is not valid to assume that the frequency distribution in
our galaxy is in steady-state.

We postpone further discussion of a uniform two-dimensional spatial distribution
to Sec.\ \ref{sec:scaledim} after presenting the setup of our numerical galactic
model.

\section{Numerical Model}\label{sec:num_model}

The second section of this paper gave a precise analytic formulation of Blandford's
argument including the extensions and improvements of Ref.\ [\onlinecite{collaboration2006}].
To understand if this argument holds in a more realistic model of our galaxy, we set up a
numerical simulation of the time evolution of a population of galactic
gravitars. This follows [\onlinecite{palomba}], using a more recently
published result [\onlinecite{hobbs}] for the initial velocity distribution of neutron
stars. To compute the spatial distribution, the equations of motion following from
the galactic potential given in Sec.\ \ref{sec:potential} are evolved over
time. The assumed initial conditions for the differential equations (i.~e.\ initial
positions and velocities of the gravitars) are described in Secs.\ \ref{sec:initialspdist}
and \ref{sec:initialvel}, respectively. Sec.\ \ref{sec:initialper} describes
the adopted distributions for the initial spin period. The results of the simulations
will be presented afterwards in Secs.\ \ref{simresult:fdist} and
\ref{sim:spdist}.

\subsection{galactic Potential and Equations of Motion}\label{sec:potential}
The motion of galactic gravitars is governed by the galactic
gravitational potential. The potential first given by Paczynski
[\onlinecite{paczynski}] is adopted. This potential describes our galaxy as
axisymmetric with respect to the rotation axis. Thus, cylindrical coordinates
$\rho$, $z$ and $\varphi$ are used.  $\rho$ denotes the distance to the
galactic rotation axis, and $z$ is the distance perpendicular to the disk.

The adopted potential represents our galaxy as composed of
three different mass components. The most massive is a nonuniform flat disk with
a radial scale of \SI{3.7}{kpc} and a $z$-direction scale of \SI{0.2}{kpc}.
The component with the second highest mass is the halo, which is described
by a density distribution $\varrho_\textsc{h} \propto \left(r^2+r_\textsc{h}^2\right)^{-1}$,
where $r_\textsc{h} = \SI{6}{kpc}$ is called the halo core radius. The central
bulge of our galaxy is represented by a spheroidal lower mass component
with a density $\varrho_\textsc{s} \propto \left(r^2+b_\textsc{s}^2\right)^{-\frac{5}{2}}$,
where $b_\textsc{s} = \SI{0.277}{kpc}$.

The corresponding potential therefore consists of three terms
\beq
\Phi\left(\rho,z\right) = \Phi_\textsc{s}\left(r\right)  + \Phi_\textsc{d}\left(\rho,z\right)  + \Phi_\textsc{h}\left(r\right).\label{potential}
\eeq
describing, respectively, the potential energy per unit mass of
the spheroid, the disk and the halo in our galaxy. The first two
components ($i=\textsc{s},\textsc{d}$) are given by
\beq
\Phi_i\left(\rho,z\right) = - GM_i \left[ \rho^2+\left(a_i + \sqrt{z^2+b_i^2}\right)^2\right]^{-\frac{1}{2}}.
\eeq
For the potential of the halo $r^2=\rho^2+z^2$ is substituted and the following
spherical symmetric expression is used
\beq
\Phi_\textsc{h}\left(r\right) = \frac{GM_\textsc{h}}{r_\textsc{h}}\left[ \frac{1}{2}\ln\left(1+\frac{r^2}{r_\textsc{h}^2}\right) + \frac{r_\textsc{h}}{r}\arctan\left(\frac{r}{r_\textsc{h}}\right)\right].
\eeq
The parameter values are shown in Table \ref{tab:pot}.
\begin{table}
\caption{Mass and scale parameters for the galactic potential}\label{tab:pot}
\begin{ruledtabular}
\begin{tabular}{rrrr}
\textsl{Disk:}&$M_\textsc{d} = \SI{8.07e10}{M_\odot}$&$a_\textsc{d} = \SI{3.7}{kpc}$&$b_\textsc{d} = \SI{0.200}{kpc}$\\
\textsl{Spheroid:}&$M_\textsc{s} = \SI{1.12e10}{M_\odot}$&$a_\textsc{s} = \SI{0}{kpc}$&$b_\textsc{s} = \SI{0.277}{kpc}$\\
\textsl{Halo:}&$M_\textsc{h} = \SI{5.00e10}{M_\odot}$&$r_\textsc{h} = \SI{6.0}{kpc}$&\\
\end{tabular}
\end{ruledtabular}
\end{table}

The axial symmetry of the galactic model
leads to
conservation of the $z$-component of the angular momentum $L_z$.
Thus, the effective potential is
\beq
\Phi_\text{eff}\left(\rho,z\right) = \Phi\left(\rho,z\right) + \frac{L_z^2}{2\rho^2}.
\eeq
The equations of motion that are evolved are
\beq
\ddot{\rho} = -\frac{\partial\Phi_\text{eff}}{\partial \rho} \quad \text{and} \quad \ddot{z} = -\frac{\partial\Phi_\text{eff}}{\partial z}. \label{eom}
\eeq
The equation of motion for $\varphi$ is given by $\rho^2\dot\varphi = L_z$.
In the simulation this equation is not used because $\varphi$ is not
evolved, but drawn from a uniform random distribution $\varphi \in \left[0,2\pi\right)$.

\subsection{Initial Spatial Distribution}\label{sec:initialspdist}
The initial spatial distribution of gravitars is proportional to the density of
massive progenitor stars of neutron stars. While there is quite good
agreement about the initial distribution in $z$-direction, the initial
distribution along the radial direction is unknown.

In the $z$-direction the initial position is drawn from a Laplacian
distribution with scale factor $z_0 = \SI{0.075}{kpc}$. The probability
of a gravitar's birth in an interval $\left[z,z+\dm\!z\right]$ is given by
\beq
p_z\left(z\right) \dm\!z = \frac{1}{2z_0}\exp\left(-\frac{\left|z\right|}{z_0}\right)\dm\!z.
\eeq

We considered three different models for the initial radial distribution
together with the given initial $z$-distribution.

The simplest radial distribution [from Ref.\ [\onlinecite{palomba}]]
is an exponential fall off with a scale factor $\rho_1 = \SI{3.2}{kpc}$.
The probability of a gravitar's birth in a distance interval
$\left[\rho,\rho+\dm\!\rho\right]$ is then 
\beq
p_1\left(\rho\right)\dm\!\rho = \frac{1}{\rho_1}\exp\left(-\frac{\rho}{\rho_1}\right) \dm\!\rho.
\eeq
Note however that this distribution leads to an extreme concentration
of neutron stars towards the galactic center.  These are not seen in pulsar surveys
[\onlinecite{YaK}].  So following [\onlinecite{paczynski}] a gamma distribution
given by
\beq
p_2\left(\rho\right)\dm\!\rho = a_\rho\frac{\rho}{\rho_2^2}\exp\left(-\frac{\rho}{\rho_2}\right)\dm\!\rho,
\eeq
is also considered, where gravitar formation in the disk is allowed for
$\rho \leqslant \SI{25}{kpc}$, and the constants are given by $\rho_2 = \SI{4.78}{kpc}$
and $a_\rho = 1.0345$. A third distribution
\beq
p_3\left(\rho\right)\dm\!\rho = \frac{\rho^5}{120\rho_3^6}\exp\left(-\frac{\rho}{\rho_3}\right)\dm\!\rho,
\eeq
with $\rho_3 = \SI{1.25}{kpc}$ taken from [\onlinecite{YaK}] is also considered. It is
fitted to the radial distribution of Population I stars which are considered to be
likely progenitors of neutron stars. Again, gravitar birth events are allowed for
$r \leqslant \SI{25}{kpc}$. On average, only one gravitar out of 14 thousand
is born with $\rho > \SI{25}{kpc}$, so the normalization constant $a_\rho \approx 1$.

\subsection{Initial Velocity}\label{sec:initialvel}
The galactic rotation determines the velocity of the supernova progenitors and
therefore also that of the newborn neutron stars.

From classical mechanics the rotational speed of a body on a circular orbit in
the axisymmetric potential (\ref{potential}) is given by
$v_\text{rot} = \sqrt{\rho\partial_\rho \Phi\left(\rho,z\right)}$.
From the initial coordinate of a gravitar the corresponding
$v_\text{rot}$ on a tangential circular orbit is  calculated (neglecting initial rotation
velocities perpendicular to the galactic disk, because all gravitars are
born with low initial values of $z$). Looking down on the Galaxy from positive
$z$-values the Galaxy is chosen to rotate counterclockwise.

Furthermore, it is assumed that gravitars are born in a supernova explosion
that will kick the newborn star. The direction of that kick is assumed to be
isotropic. The kick speed is drawn from a Maxwellian distribution
with a mean velocity $\overline{v} = \SI{430}{km/s}$. The probability for the kick
speed to be in an interval $\left[v_\text{kick}, v_\text{kick}+\dm\!v_\text{kick}\right]$ is
\beq p_v\left(v_\text{kick}\right) \dm\!v_\text{kick} = \frac{32v_\text{kick}^2}{\pi^2 \overline{v}^3}\exp\left( -\frac{4v_\text{kick}^2}{\pi\overline{v}^2} \right)\dm\!v_\text{kick}.\eeq
It is assumed that this distribution from Ref.\ [\onlinecite{hobbs}] may be used
for gravitars as well as for pulsars.

\subsection{Initial Period Distribution}\label{sec:initialper}
Because of the conservation of angular momentum in the supernova
event and the much smaller radius of the gravitar compared with its
progenitor, newborn neutron stars will spin rapidly. 

We considered three models for the distribution of the initial rotation
periods following the models given in Ref.\ [\onlinecite{palomba}].
The existence of three different models reflects our ignorance of
the actual distribution of initial periods.

The first model is described by a lognormal distribution
\beq
p_{P_0}\left(P_0\right) = \frac{1}{\sqrt{2\pi}\sigma P_0}\exp\left[ -\frac{1}{2\sigma^2}\left( \ln\left(P_0\right) - \ln\left(\overline{P_0}\right) \right)^2 \right] \label{eq:P0}
\eeq
where $P_0$ is measured in seconds and where the values $\sigma = 0.69$
and $\overline{P_0} = \SI{5}{ms}$ are taken from Ref.\ [\onlinecite{arzoumanian}].
Gravitars with $P_0 < \SI{0.5}{ms}$ are excluded.

The second model is using the same probability distribution as
the first model but every initial period $P_0 < \SI{10}{ms}$ is set
to \SI{10}{ms} exactly. In this way the possible presence of
r-modes in young neutron stars is mimicked. These modes can
dissipate rotational energy of the newborn neutron star and
increase its initial period to about \SI{10}{ms} within \SI{1}{yr}.

The third model considered is a further extension of the
second one. It includes the effects of matter fall-back after the
supernova explosion. The increase of angular momentum by
the accreting matter could counteract the r-mode induced
deceleration. The resulting initial period will approach an
intermediate value. The choice from [\onlinecite{palomba}]
to draw the initial period from a uniform distribution between
\SI{2}{ms} and \SI{15}{ms} is adopted

\subsection{Coding and Implementation}\label{sec:coding}

With the initial distributions from Secs.\ \ref{sec:initialspdist} and
\ref{sec:initialvel} and the equations of motion (\ref{eom}) at our disposal,
it is a straightforward problem to find the spatial distribution of a
population of gravitars at the present time.

The code for the simulation is written in C. The equations of motion (\ref{eom})
are integrated via a Burlisch-Stoer method in combination
with Stoermer's rule for the direct discretization of a system of second-order differential equations using routines described in [\onlinecite{NR}].
Over the integration time the total energy is conserved to one part in
\num{e6}.

For the derivation of the frequency distributions (\ref{f:dist}) the probability
distributions are implemented by random number generators
and functions from the GNU Scientific Library (GSL) [\onlinecite{gsl}].
The fractional frequency density (\ref{ffracdens}) is obtained via a Monte-Carlo
integration using \num{2e12} random values of initial frequency and
a uniform distribution of ages.

Depending on the model of the initial spatial distribution the integration
of \num{e6} neutron star trajectories over a time of \SI{200}{Myrs} takes
\SI{2.5}{mins} to \SI{13.3}{mins} on an AMD Opteron 185 processor. Most
of the simulations were done on the Morgane cluster at the AEI in Potsdam.

\section{Results}\label{sec:results}

\subsection{Frequency Distributions}\label{simresult:fdist}
In the derivation of the generalized result (\ref{eq:M_general}) the assumption of a steady-state
frequency distribution (\ref{ffracdens_simple}) is (along with the two-dimensional
uniform spatial distribution) the key to the independence of the maximum
amplitude on ellipticity and frequency. Let us therefore first have a look
at the frequency distributions that result from a continuous distribution
of initial frequencies and compare them with the corresponding density
resulting from a single birth frequency (\ref{ffracdens_simple}).

\begin{figure}
\includegraphics[width=\columnwidth]{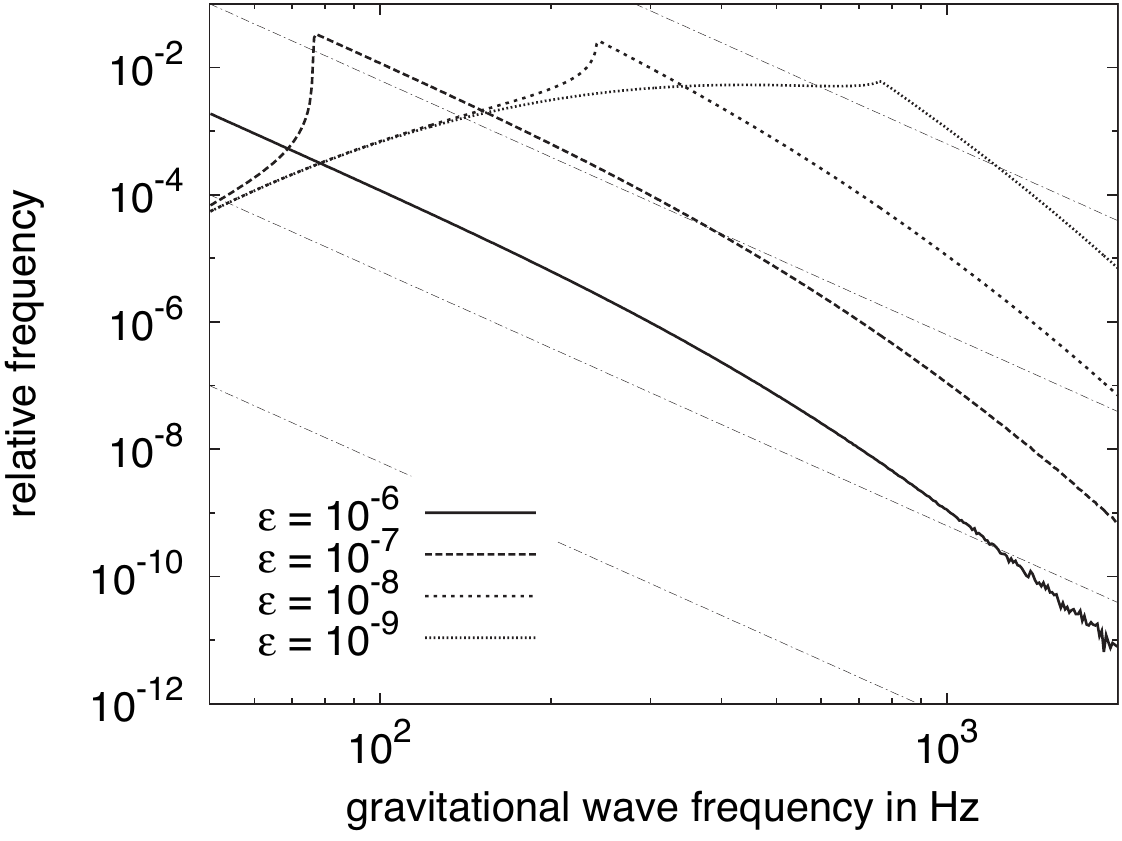}
\caption{The distribution $\hat\varrho_f\left(\varepsilon,f\right)$
in frequency after $\overline{t}=\SI{13.6}{Gyrs}$ for varying
ellipticity using a lognormal distribution of initial periods in the
range $f\in\left[\SI{50}{Hz},\SI{2000}{Hz}\right]$. The dashed
lines correspond to a slope $f^{-4}$, as Eq.\ (\ref{ffracdens_simple})
would predict for a single high birth frequency. The kink
in the graphs for $\varepsilon \lesssim \num{e-7}$ is at the frequency
given by Eq.\ (\ref{eps_f}).}\label{n_f_t2}
\end{figure}

Fig.\ \ref{n_f_t2} shows the results of the Monte Carlo integration for
$\overline{t}=\SI{13.6}{Gyrs}$ using a lognormal distribution of initial
periods and a fixed ellipticity for each run in the frequency range
$f\in\left[\SI{50}{Hz},\SI{2000}{Hz}\right]$. The ellipticity varies over 3
orders of magnitude from \num{e-9} to \num{e-6}. The dashed lines
correspond to a scaling proportional to $f^{-4}$, which results from a single birth
frequency and shows a steady-state distribution as calculated in Eq.\ 
(\ref{ffracdens_simple}).

For $\varepsilon=\num{e-6}$ the population is close to a steady state
at the present time because the scaling is nearly  proportional to $f^{-4}$.
For smaller ellipticities one can identify a kink in the density function at
a frequency $\tilde f$ as given by (\ref{eps_f}). The kink is due to gravitars born
at high frequencies that are too young to have spun down to lower
frequencies. The frequency distribution for $\varepsilon \lesssim \num{e-7}$
is not in a steady state in the frequency range of highest sensitivity for modern
interferometric detectors, which is between \SI{100}{Hz} and \SI{300}{Hz}.

We also note that $\hat\varrho_f\left(\varepsilon,f\right)$ does not scale as
$\varepsilon^{-2}$ in all frequency bands. It only scales as $\varepsilon^{-2}$
at high frequencies.

\begin{figure}
\includegraphics[width=\columnwidth]{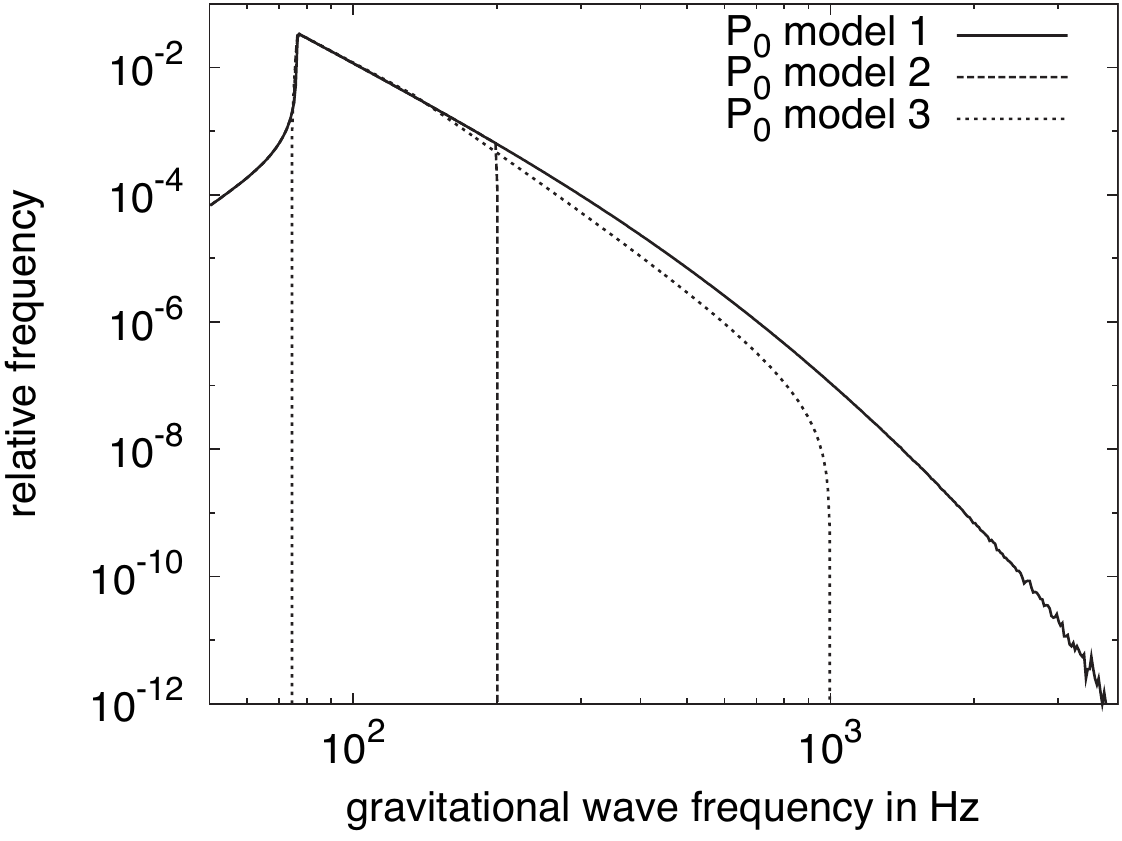}
\caption{The distribution in frequency space in the range
$f\in\left[\SI{50}{Hz},\SI{4000}{Hz}\right]$ after $\overline{t}=\SI{13.6}{Gyrs}$
for $\varepsilon = \num{e-7}$ for three different models of the initial gravitar
frequency distribution.}\label{modeldependence}
\end{figure}

Let us now compare the fractional frequency densities that result from
different models of initial frequency distributions for a fixed ellipticity. Fig.\ 
\ref{modeldependence} shows the distribution in frequency space after
an evolution time of $\overline{t}= \SI{13.6}{Gyrs}$ for $\varepsilon = \num{e-7}$.
Each graph corresponds to one of the models for the initial frequency distribution.

If the evolution time is long compared with the spin-down time
$\tau_\textsc{gw}$ (\ref{sdtimescale}), most of the gravitars will
have spun down to low frequencies, and the distribution will
be dominated by those older sources. Yet, if the respective model
has upper or lower limits on the birth frequencies one cannot
expect the fractional densities to agree near these boundaries.
However, in frequency bands of interest for modern interferometric detectors
(\SI{100}{Hz} to \SI{300}{Hz}) the distributions only show minor differences
between the different models.

\subsection{Spatial Steady-State Distribution and Timescales}\label{sim:spdist}
Since the gravitars are born in a thin disk and receive an isotropic kick by the supernova,
they tend to leave the disk after some Myrs.  They either escape the galactic gravitational
potential, or are bound to the Galaxy on some ``orbit''. The numerical simulation is used
to find the timescale on which these  processes wash out the imprint of the initial spatial
distribution.

Let us introduce the function $\hat M\left(r,t\right)$, which is the number of
gravitars in a ball of radius $r$ around the position of the Sun that were formed
a time $t$ ago. The radial probability distribution $\varrho_\text{r}\left(r,t\right)$
as introduced in Sec.\ \ref{restate:bland} is related to $\hat M$ via the derivative
with respect to $r$
\beq
\varrho_\text{r}\left(r,t\right) \dm\!r = \frac{1}{N_\text{tot}}\partial_r \hat M\left(r,t\right) \dm\!r.
\eeq

To obtain a dynamical picture, $\hat M\left(r,t\right)$ is computed in steps of  \SI{1}{Myr}
from \SI{0}{Myrs} to \SI{200}{Myrs}.  For each of the 201 values of integration time the
trajectories of $N_\text{tot} = \num{e9}$ galactic gravitars are evolved using the
numerical integration methods described in Sec.\ \ref{sec:coding} and the
radial distance distribution is derived from their final positions. The radial resolution
is chosen as \SI{2.5}{pc} for $\SI{0}{kpc} \leqslant r < \SI{12}{kpc}$ and as \SI{100}{pc} for
$\SI{12}{kpc} \leqslant r \leqslant \SI{20}{kpc}$

\begin{figure}
\includegraphics[width=\columnwidth]{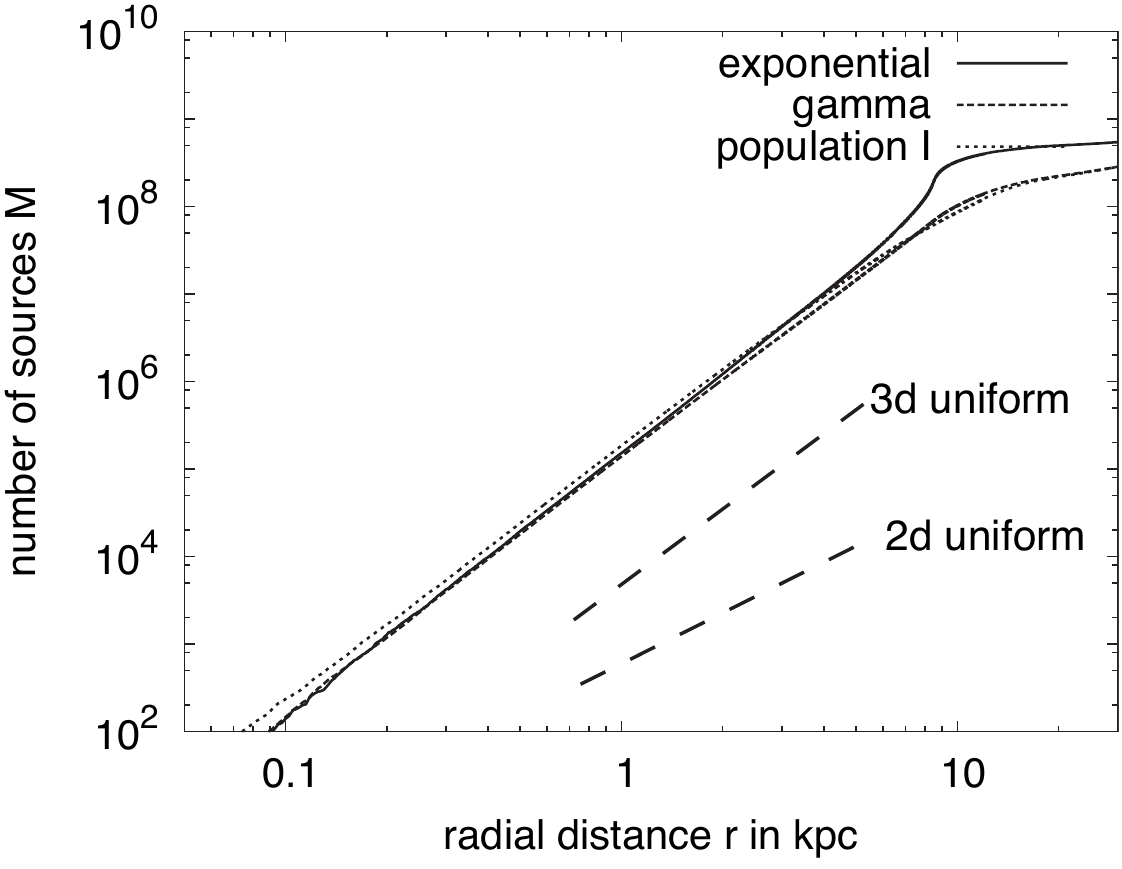}
\caption{The distribution of gravitars as a function of radial distance from the
Sun. The graphs show the number of gravitars inside spheres of radius $r$
plotted against $r$. The age of all sources is $t= \SI{200}{Myrs}$. The scaling
dimension of these graphs is shown in Fig.\ \ref{local_dim}. The straight
(dashed) lines show the corresponding slope for a uniform two- (three-) dimensional
distribution.}\label{dist_scaling_log}
\end{figure}

Fig.\ \ref{dist_scaling_log} shows the number of gravitars inside balls of radius $r$
around the position of the Sun, which were formed $t=\SI{200}{Myrs}$ ago, for the
different initial radial distributions. The saturation near $r = \SI{8.5}{kpc}$ is due
to the high density of gravitars near the galactic center.

 The simulations show that the spatial distribution settles into a state of
 equilibrium for $t \approx \SI{200}{Myrs}$. There is no significant difference
 between the distributions for $t = \SI{200}{Myrs}$ and $t = \SI{2}{Gyrs}$. By the
 age of \SI{200}{Myrs} the initial distribution is washed out;
 evolution over longer times no longer changes the radial distribution $\hat M\left(r\right)$.

\subsection{Scaling Dimension of the Spatial Distribution}\label{sec:scaledim}
The assumption of  a two-dimensional and uniform spatial distribution
of gravitars at the present time is crucial for Blandford's argument. 
The numerical simulation can test if these assumptions are valid or not.

A useful concept is that of the
\textsl{scaling dimension}. To obtain the scaling dimension, the
function $\hat M\left(r\right)$ as introduced in the previous section is used.
Assume a uniform distribution, and describe the number
of sources inside each ball as a function of its radius $r$ by a
simple power law
\beq
\hat M\left(r\right) \propto r^{D}.\label{scalingdim:theoret}
\eeq
$D$ is called the \textsl{scaling dimension} of the distribution.

Even an exactly two-dimensional spatial distribution of gravitars (e.~g.\ the galactic disk)
can effectively scale as a $D$-dimensional object due to density gradients. Note, that the
scaling dimension is a local quantity depending on the position of evaluation. 

\begin{figure}
\includegraphics[width=\columnwidth]{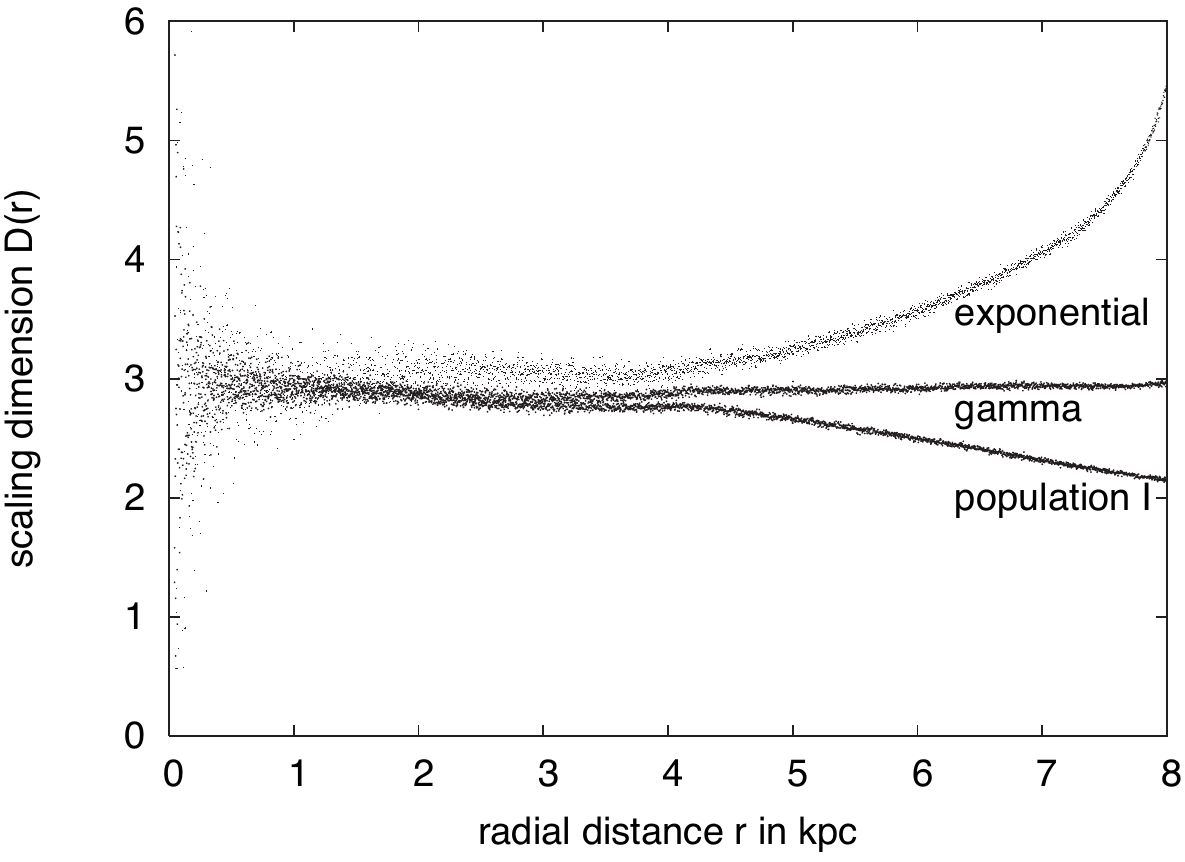}
\caption{The local scaling dimension $D\left(r\right)$ of the spatial distribution
of gravitars around the position of the Sun as calculated by (\ref{eq:dofr}). The
corresponding radial distributions of sources are shown in Fig.\ 
\ref{dist_scaling_log}. Note, that here the graphs are only shown for
$r\leqslant\SI{8}{kpc}$.}\label{local_dim}
\end{figure}

To illustrate the scaling properties of the evolved galactic spatial distribution of gravitars,
the local scaling dimension is computed for $r \leqslant \SI{8}{kpc}$ and is
shown in Fig.\ \ref{local_dim}.
From (\ref{scalingdim:theoret}) the scaling dimension $D\left(r\right)$ can be derived via
\beq
D\left(r\right) = \frac{r\partial_r \hat M\left(r\right)}{\hat M\left(r\right)}\label{eq:dofr}
\eeq
The differentiation is computed numerically using a cubic splining on the
tabulated values.

The scattering of points for $r \lesssim \SI{1}{kpc}$ is due to the small number
of sources at short distances and resulting numerical noise.

The scaling dimension at every distance to the Sun is greater than 2. For the
first model of initial radial distributions it even reaches values $D \geqslant 5$.
For the second model there is a slight increase in the scaling dimension towards
the galactic center where the scaling dimension $D\approx3$. For the last model
the scaling dimension decreases with larger radial distance but always $D > 2$.

We conclude that the scaling dimension of the population of gravitars in the
model of our galaxy is significantly larger than 2. More precisely
averaging $D$ over distances $r\leqslant \SI{2}{kpc}$ -- where $D$ is nearly
constant and independent of the initial radial distribution -- yields $D \approx 2.95$.
 
\section{The Strongest Continuous Gravitational-Wave Signal}\label{bland:revise}
Using the frequency and spatial distributions obtained from our
galactic simulation, it is straightforward to derive the maximum
expected amplitude of continuous gravitational waves from gravitars.

\subsection{Numerical Method}

Let us first describe the numerical method for computing the maximum
expected amplitude of the gravitational waves using the distributions in space 
and frequency as presented in Secs.\ \ref{simresult:fdist} and \ref{sim:spdist}.

Eq.\ (\ref{eq:M}) is used to obtain the value of $M\left(f_1,ef_1,h_\text{max}\right)$
for a given frequency band $\left[f_1,e f_1\right]$ ($\ln\left(e\right) = 1$) and a trial
value of $h_\text{max}$. To compare the results with Ref.\ 
[\onlinecite{collaboration2006}] $h_\text{max}$ is tuned via a bisection method
within $\pm 2.5\%$ to the target value $M_\text{tar}=0.5$ such that $0.4875 \leqslant
M\left(f_1,ef_1,h_\text{max}\right) \leqslant 0.5125$.

As described in Sec.\ \ref{sim:spdist} the function $\hat M\left(r,t\right)$ giving the
number of gravitars with a solar radial distance less than $r$ is tabulated. From
these values $\varrho_\text{r}\left(r,t\right) = \partial_r \hat M\left(r,t\right)/N_\text{tot}$
is numerically computed via a cubic splining method.

The distribution in frequency space $\varrho_f\left(\varepsilon,f,t\right)$ is taken from
Eq.\ (\ref{f:dist}) with distributions of initial periods as described in Sec.\ 
\ref{sec:initialper}.

Given a frequency band a high value of $h_\text{max}$ is chosen as trial value.
Then the frequency is chosen fixed at the lower boundary of the band and the
integration over $h\in\left[h_\text{max},h_\text{up}\right]$ in (\ref{eq:M})
is conducted by calculating from (\ref{strain}) the corresponding $r\left(h\right)$
and inserting into the interpolated $m\left(r,t\right)$. For the upper limit $h_\text{up}$
of the integration the value that would obtained if the gravitar was at the closest
possible distance is taken. Then no contribution to the integral is lost. After integrating
over $h$ the integration over frequency in the chosen band $\left[f_1,ef_1\right]$ 
is done and weighted by $\varrho_f\left(\varepsilon,f,t\right)$.

The result of these two integrations is then integrated over all times
$t \in\left[0, \overline{t} = \SI{13.6}{Gyrs}\right]$ where the timestep between
two evaluations is chosen as $\dm\!t = \min\left\{\SI{1}{Myr},\frac{\tau_\textsc{gw}}{10}\right\}$
to obtain a sufficiently fine timestep to track both spatial and frequency evolution.
This integration over time is weighted by a constant birthrate of
$\dot n\left(t\right) = \left(\SI{30}{yrs}\right)^{-1}$.

\subsection{Maximum Expected Amplitude}\label{max_exp_amp}

Using the results of the simulation, one can see if the value of $h_\text{max}$
(obtained by the method described in the previous section) differs from the
one predicted by Blandford's result as extended and improved in Ref.\ 
[\onlinecite{collaboration2006}]. 

To avoid boundary effects, from now only the first model for the
distribution of initial frequencies is considered. In all frequency
bands, the other models always give a smaller maximum amplitude of
gravitational waves $h_\text{max}$.

\begin{figure}
\includegraphics[width=\columnwidth]{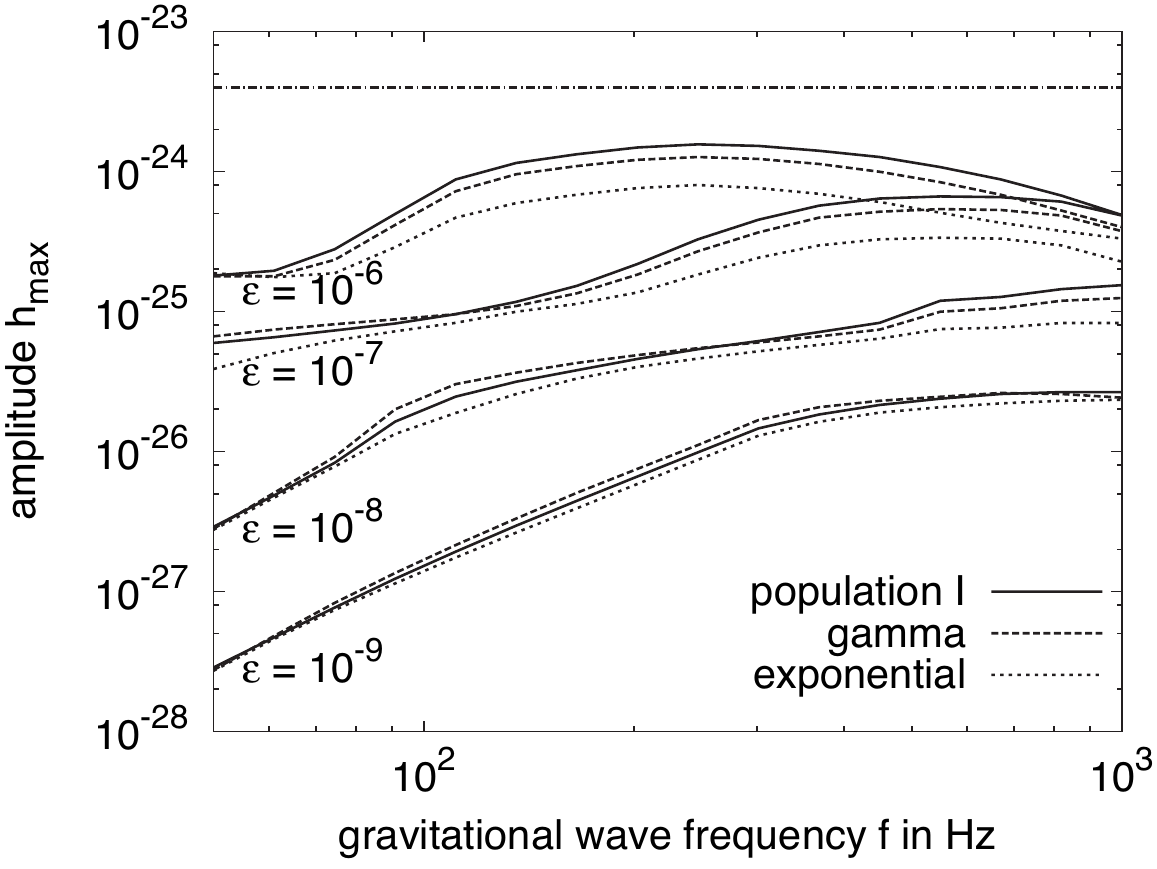}
\caption{\label{amplitude} The maximum strain amplitude of gravitational
  waves $h_\text{max}$ from galactic gravitars in frequency bands
  $\left[f,ef\right]$ in the range
  $f\in\left[\SI{50}{Hz},\SI{1000}{Hz}\right]$. Each plotted point
  $h_\text{max}\left(f\right)$ is the value $h_\text{max}$ calculated for
  a frequency band $\left[f,ef\right]$ which is one
  natural-logarithmic-octave wide.  The three curves show different initial spatial distribution models.
  For contrast, the dotted, straight line
  (independent of frequency and ellipticity $\varepsilon$) shows the value
  of $h_\text{max}$ from [\onlinecite{collaboration2006}], which
  improved and extended Blandford's argument.}
\end{figure}

Fig.\ \ref{amplitude} shows the resulting value of $h_\text{max}$ for different
values of the ellipticity and all spatial distribution models. The graphs are to
be understood as follows: the maximum amplitude of gravitational waves is
calculated as described in the previous section for every choice of initial
spatial distributions and for frequency bands $\left[f,ef\right]$. The graphs
shows the value $h_\text{max}$ obtained in such a band as a single point at
$\left(f,h_\text{max}\right)$.

\begin{figure}
\includegraphics[width=\columnwidth]{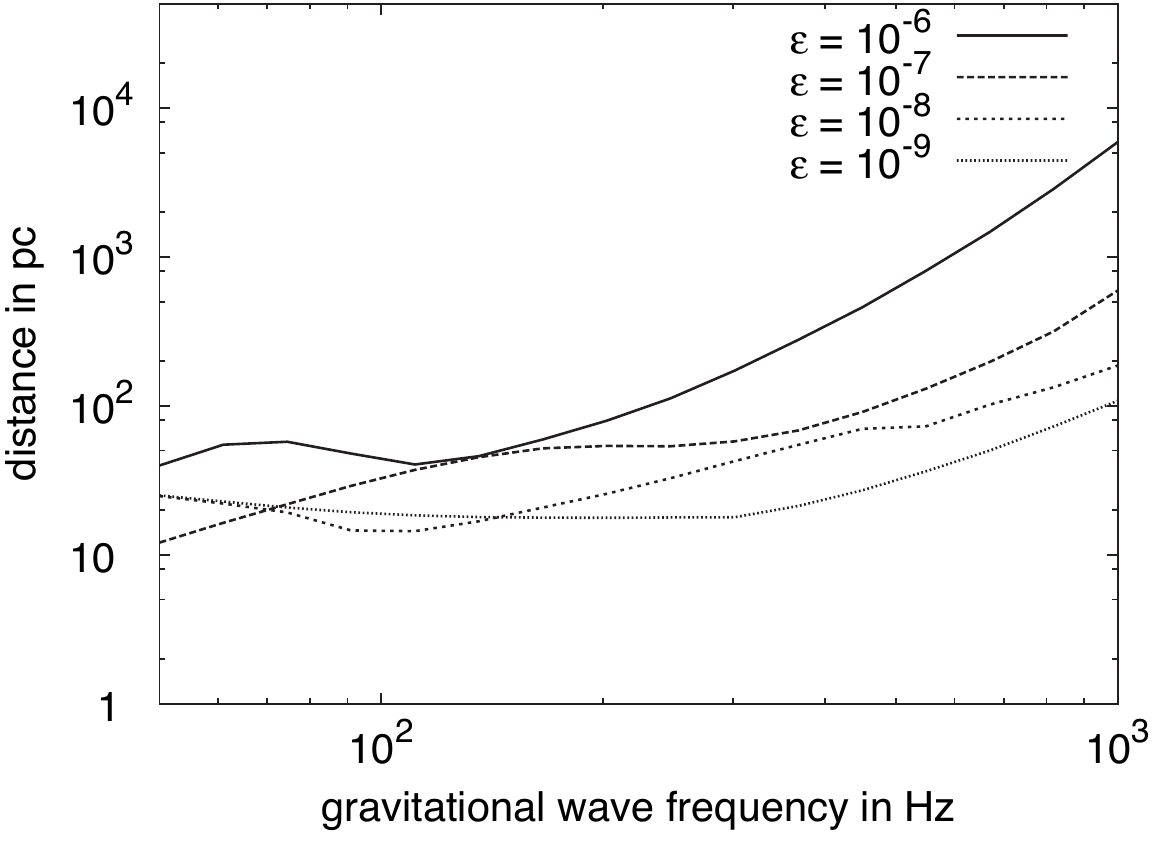}
\caption{\label{distanceplot}
The expected distance to the strongest gravitar, corresponding to Fig.\ \ref{amplitude}.
Because each frequency axis value of Fig.~\ref{amplitude} refers to an entire logarithmic
octave $\ln\left(f_2/f_1\right) = 1$ range, in principle the plots should show
a range of distances which is two logarithmic octave wide: $\ln\left(h_2/h_1\right) = 2$.
To simplify the appearance of this plot, we show only a single curve at the
central value. The expected range of distances ranges from a factor of $1/e$
below this plot to $e$ above this plot.}
\end{figure}

From Fig.\ \ref{amplitude} it is obvious that the assumptions of Blandford's
argument are not fulfilled for a realistic model of our galaxy. The graphs for different
values of ellipticity do not line up, and each single graph is curved. Thus, the
maximum amplitude of gravitational waves from galactic gravitars
does depend on both the ellipticity and frequency.

For highly deformed gravitars the graphs are nearly flat indicating a
weak dependence on $f$, while for low values of $\varepsilon$ the
previously-discussed kink appears.

The choice of the initial radial distribution causes only small differences. The
inter-model differences are usually of the order of 10\%, and in the worst case are about 50\%
for a particular ellipticity and frequency. This is because the strongest
gravitational waves are emitted by gravitars at very small distances \footnote{E.~g.\ a
gravitar with $f=\SI{2}{kHz}$, $\varepsilon = \num{e-6}$must be as close as
$r = \SI{1}{kpc}$.}. As can be seen from Fig.\ \ref{dist_scaling_log},
the distribution for radial distances $r\lesssim\SI{5}{kpc}$ from the Sun is nearly the
same independent of the initial radial distribution.

Assuming the highest possible ellipticity $\varepsilon = \num{e-6}$ for the
gravitars, the strongest signal has an amplitude of $h_\text{max} \approx
\num{1.6e-24}$ in the band $\left[\SI{250}{Hz},\SI{680}{Hz}\right]$ which is
improved (strengthened) by a factor of $\approx3$ compared with the value given
in [\onlinecite{collaboration2006}]. Note that this improvement factor would be
smaller if larger values of $\varepsilon$ were considered.  In any case, the
improvement factor is substantially larger at other frequencies.

In the case of the more realistic value $\varepsilon = \num{e-7}$, the estimate is
$h_\text{max} \approx \num{6.6e-25}$ in the band $\left[\SI{550}{Hz},\SI{1500}{Hz}
\right]$ and is lower than the simple analytic estimate by a factor of 6.

Table \ref{table:hmax} lists the maximum values for the amplitude of the
strongest gravitational waves for all adopted values of the ellipticity and the
frequency band in which the maximum amplitude is expected. 
\begin{table}
\caption{Maximum values for the amplitude $h_\text{max}$ of gravitational
waves in dependence on the ellipticity $\varepsilon$.}\label{table:hmax}
\begin{ruledtabular}
\begin{tabular}{ccc}
$\varepsilon$&$h_\text{max}$&frequency band\\
\num{e-6} & \num{1.6e-24} & $\left[\SI{250}{Hz},\SI{680}{Hz}\right]$\\
\num{e-7} & \num{6.6e-25} & $\left[\SI{550}{Hz},\SI{1500}{Hz}\right]$\\
\num{e-8} & \num{1.5e-25} & $\left[\SI{1000}{Hz},\SI{2800}{Hz}\right]$\\
\num{e-9} & \num{2.7e-26} & $\left[\SI{1000}{Hz},\SI{2800}{Hz}\right]$\\
\end{tabular}
\end{ruledtabular}
\end{table}

To illustrate the expected distance to the strongest gravitars we used the results of Fig.\ \ref{amplitude}
together with Eq.\ (\ref{strain}) to compute the distance to these sources as a function of the 
gravitational-wave frequency interval and the ellipticity. The result is shown in Fig.\ \ref{distanceplot}

We conclude that the assumptions of Blandford's argument do not hold
in our galactic model. The expected gravitar spatial distribution is
not a two-dimensional uniform thin disk, and the expected gravitar
frequency distribution is not yet in a steady state for realistic values
of neutron star ellipticity.

\subsection{Remarks on an Upper Limit}
In the previous section we obtained the expected maximum amplitude of a
gravitational-wave signal from a deformed neutron star spinning down purely by
gravitational waves. Let us now address the question whether this value poses
an upper limit on the gravitational-wave amplitude from objects that spin down
partly by gravitational waves and partly by electromagnetic dipolar emission.
It must be stressed that recycled millisecond pulsars are not covered here,
since spin-up is {\it not} considered. 

Ref.\ [\onlinecite{collaboration2006}] gave a clever argument about why the expected
maximum gravitational-wave strain from gravitars sets an upper limit for all
neutron stars that have not gone through an accretion-powered spin-up phase.
However that argument implicitly assumes only birth frequencies {\it above} the
observed frequency band for the pulsars, and also assumes that the pulsar
population frequency-space distribution is in steady-state in the observed
frequency band. Here, we do not make either of these assumptions.

In this section we show that the results from Fig.\ \ref{amplitude} are a
strict upper limit for frequencies $f>\SI{250}{Hz}$. This is related to the distribution
of initial frequencies which has its  maximum at $f_0=\SI{250}{Hz}$ and is
decreasing monotonically for larger values of initial frequency. To maximize the
number of sources in a frequency band at a given time in this regime, the slowest
possible spin-down is required. Any faster spin-down only would remove sources
from this frequency band without adding more new ones from higher frequencies. Since
the slowest possible spin-down without weakening the gravitational-wave signal
is attained by switching off dipolar emission, the results from Fig.\ \ref{amplitude}
are a strict upper limit on the gravitational-wave amplitude for $f>\SI{250}{Hz}$.

To formulate a rigorous proof of this claim, we characterize the
generalized spin-down such a ``mixed'' neutron star will experience
with a spin-down parameter $\hat y$. The spin-down from electromagnetic
dipolar emission in terms of the rotation frequency $\nu$ is given by
\beq
\dot \nu_\text{dip} = -\frac{2\pi^2}{3c^3} \frac{B_\text{p}^2R^6\sin^2\left(\alpha\right)}{I}\nu^3
=: \gamma_\text{dip}\nu^3,
\eeq
where $B_\text{p}$ is the magnetic field strength at the neutron star's magnetic pole,
$\alpha$ is the angle between the rotation axis and the magnetic field, and
$R$ is the radius of the neutron star. Rewriting Eq.\ (\ref{dotf}) the spin-down
from gravitational waves is
\beq
\dot \nu_\text{gw} = -\frac{512\pi^4G}{5c^5} I\varepsilon^2\nu^5
=: \gamma_\text{gw}\nu^5.
\eeq
A neutron star emitting energy by both mechanisms at once will experience a total
spin-down
\beq
\dot \nu = \gamma_\text{gw}\nu^5 + \gamma_\text{dip}\nu^3.
\eeq
This differential equation cannot (in contrast to (\ref{dotf})) be solved analytically
for $\nu\left(t\right)$, yet it can be integrated to give the time $t\left(\nu,\nu_0\right)$
in which a neutron star spins down from rotation frequency $\nu_0$ to $\nu$:
\beq
t\left(\nu,\nu_0\right) =\frac{1}{2\left|\gamma_\text{dip}\right|}
\left[ \frac{\nu_0^2 - \nu^2}{\nu_0^2\nu^2} +
\hat y \ln\left(\frac{\nu^2}{\nu_0^2}{\left( \frac{1+ \hat y\nu_0^2} {1+ \hat y\nu^2} \right)} \right)\right],\label{t_general}
\eeq
where $\hat y := \gamma_\text{gw}/\gamma_\text{dip}$ is the \textsl{general spin-down
parameter}. Note, that $\hat y \rightarrow 0$ corresponds to switching off gravitational
wave emission, while $\hat y \rightarrow \infty$ leads to pure gravitational-wave spin-down.
Taking these limits in (\ref{t_general}), one easily recovers the equations for pure
dipolar and gravitational-wave spin-down, respectively, which can be solved analytically
for $\nu\left(t\right)$.

Let us now turn to the derivation of the expected maximum gravitational wave
amplitude from such neutron stars. The derivation given in Sec.\ \ref{restate:bland}
is straightforwardly modified to incorporate the generalized spin-down. The
generalization of the present-time gravitational-wave frequency distribution $\varrho_f$
is obtained by writing Eq.\ (\ref{f:dist_first}) for fixed $t$ and $\hat y$ as
\beq
\varrho_f\left(f, t, \hat y\right)\dm\!f  =  \varrho_{f_0}\left(f_0\left(f, t, \hat y\right)\right)\frac{\partial f_0\left(f, t, \hat y\right)}{\partial f}\dm\!f \label{rho_f_haty}.
\eeq
Note, that (\ref{t_general}) cannot be solved analytically for $f_0\left(f,t,\hat y\right)
= 2\nu_0\left(f,t,\hat y\right)$. Care has to be taken to evaluate the partial derivative.
Taking the total derivative of $t\left(\nu,\nu_0\right) = \text{const}$ with respect to $\nu$ and
application of the chain rule yields by a straightforward calculation
\beq
\frac{\partial f_0}{\partial f} = \frac{\partial \nu_0}{\partial \nu} = -\frac{\partial t}{\partial \nu}\cdot\left(\frac{\partial t}{\partial \nu_0}\right)^{-1}.
\eeq
Evaluation of this expression by use of (\ref{t_general}) finally leads to
\beq
\frac{\partial f_0}{\partial f} = \frac{f_0^3\left(4+\hat y f_0^2\right)}{f^3\left(4+\hat y f^2\right)}.\label{df0bydf}
\eeq
It is easy to see that $\hat y \rightarrow \infty$ implies $\frac{\partial f_0}{\partial f}
\rightarrow \frac{f_0^5}{f^5}$, reproducing Eq.\ (\ref{f:dist}).

However, there is no conceptual difference between a spin-down governed
by $\varepsilon$ and one governed by $\hat y$. One can write Eq.\ (\ref{eq:M})
giving the number of neutron stars with fixed spin-down parameter $\hat y$ in a frequency
band $\left[f_1,f_2\right]$ and gravitational-wave amplitude $h\geqslant h_\text{max}$ as
\begin{align}
&M\left(f_1,f_2,h_\text{max},\hat y\right) =\nonumber\\
&\int_0^{\overline{t}}\!\!\dm\!t\,\dot n\left(t\right) \int_{f_1}^{f_2} \dm\!f
\varrho_f\left(f,t,\hat y\right)\int_{h_\text{max}}^{\infty}\dm\!h\,
\varrho_\text{r}\left(r\left(h\right),t\right)\frac{\dm r\left(h\right)}{\dm\!h}.\label{eq:M_hat_y}
\end{align}

The fact that $h_\text{max}$ for given $\hat y$, $f_1$, $f_2$ and $M$ is an upper limit on
the gravitational-wave amplitude from a population of neutron stars, whose spin-down
is governed by $\hat y$, can be rephrased as follows:
$h_\text{max}$ is an upper limit, if $M\left(f_1,f_2,h_\text{max},\hat y\right)$ is maximal as
a function of $\hat y$. If $M$ is not maximal as a function of $\hat y$, then a larger value
of $h_\text{max}$ in the same frequency band could be found for a different value of
$\hat y$ giving the same $M$. Thus, it is necessary to identify maxima of
$M\left(f_1,f_2,h_\text{max},\hat y\right)$ in $\hat y$.  These satisfy
\beq
\frac{\dm}{\dm \hat y} M\left(f_1,f_2,h_\text{max},\hat y\right) = 0.
\eeq
From Eq.\ (\ref{eq:M_hat_y}) it is clear, that the only term affected by the
derivative is $\varrho_f\left(f,t,\hat y\right)$. Applying the chain rule to
(\ref{rho_f_haty}) after inserting (\ref{df0bydf}) one obtains
\beq
\frac{\dm}{\dm \hat y} \varrho_f = \frac{\partial\varrho_{f_0}}{\partial f_0}\cdot
\frac{\partial f_0}{\partial \hat y}\cdot\frac{f_0^3\left(4+\hat y f_0^2\right)}{f^3\left(4+\hat y f^2\right)} +
\varrho_{f_0}\cdot\frac{f_0^3\left(f_0^2 - f^2\right)}{f^3\left(4+\hat y f^2\right)^2}\label{drho_fbydy},
\eeq
where the arguments of the functions are suppressed.

Let us show that the last equation implies that the maximum expected amplitude
as given in Sec.\ \ref{max_exp_amp} is a rigorous upper limit for frequencies
$f_0$ for which
\beq
\frac{\partial \varrho_{f_0}}{\partial_{f_0}} \left|_{f_0\left(f,t,\hat y\right)} \right. \leqslant 0 \; \forall \; f,t.\label{uplimit_condition}
\eeq
The second summand on the right-hand side of (\ref{drho_fbydy}) is positive
$\forall \, f, t,\hat y$. The last factor in the first summand is always positive. It is also clear, that
$\partial_{\hat y}f_0 < 0  \; \forall \; f,t$ for a fixed value of $\varepsilon$ \footnote{Increasing
$\hat y$ for fixed $\varepsilon$ and therefore fixed $\gamma_\text{gw}$ is only possible by
decreasing $\gamma_\text{dip}$, that is reducing the amount of energy radiated away by
dipolar spin-down. Accordingly, the total spin-down is slower and $f_0$ less.}.
Therefore, if (\ref{uplimit_condition}) holds true, it follows $\frac{\dm}{\dm\hat y}
\varrho_f > 0 \; \forall \; f,t,\hat y$ and $\frac{\dm}{\dm \hat y} M > 0 \; \forall \; \hat y$.
Thus, if (\ref{uplimit_condition}) is fulfilled, the global maximum of $M$ as a function
of $\hat y$ is reached for $\hat y \rightarrow \infty$, that is for pure gravitational wave
spin-down.

As mentioned earlier, for the model distribution of initial frequencies adopted to produce
Fig.\ \ref{amplitude} has its maximum at $f_0 = \SI{250}{Hz}$ and is monotonically
decreasing for larger values of initial frequencies. Therefore, Eq.\ (\ref{uplimit_condition})
is fulfilled for $f\geqslant\SI{250}{Hz}$ in Fig.\ \ref{amplitude}, and the graphs shown there
are a strict upper limit on the gravitational-wave amplitude from neutron stars spinning down
by gravitational waves and electromagnetic dipolar emission for $f>\SI{250}{Hz}$.

\section{Conclusions \label{s:conclusion}}
We have used analytical arguments and the results of a numerical simulation
to show that the assumptions of Blandford's argument do not hold in a realistic model of
our galaxy.

The assumptions (both in the original and in revised formulations of the argument)
cannot be fulfilled for realistic values of ellipticity. The spatial scaling dimension
$D$ of an evolved neutron star distribution fulfills $D > 2$  making a simple two-dimensional
model invalid. The distribution in frequency will not be in a steady state at the present
time for realistic values of ellipticity.

Because these two assumptions do not hold, the
simple geometrical reasoning behind Blandford's argument is not valid.
The numerical simulations provide an improved estimate of the expected
maximum amplitude of gravitational waves from gravitars. We also showed that
for frequencies $f>\SI{250}{Hz}$ this maximum amplitude is an upper limit for
gravitational waves from neutron stars that spin down by gravitational waves and
electromagnetic dipole emission. 

Although the expected maximum amplitude is lower by about 1 order of magnitude
compared to the previous estimates, we would like to stress that in all of the models so
far the influence of the Gould belt has been neglected. This young star-forming region
(age $\sim \SI{40}{Myrs}$) near the Sun is characterized by an abundance of massive
O- and B-type stars enriching the solar neighborhood with young neutron stars. If there
exists a population of gravitars born in the Gould belt, their gravitational-wave signals
are more likely to be the first ones to be detected.

\section{Acknowledgments}
We thank Cristiano Palomba for useful discussions and for helping to
compare the results of his numerical simulations with our own work,
and Curt Cutler for helpful discussions and calling our attention to
the nice upper-limit argument given in Ref.\ 
[\onlinecite{collaboration2006}].  We also thank the referee, for some
suggested rewording and clarification of the status of the gravitar
existence arguments.  B.~K.\ thanks the IMPRS on Gravitational Wave
Astronomy for its support.  This work was supported in part by DFG
Grant SFB/Transregio 7 ``Gravitational Wave Astronomy''.

\bibliography{blandford}

\end{document}